\documentclass{ws-procs9x6}
\def\etal {{\it et al.}}
\begin{document}
%\vspace*{-2.5cm}

\title{TESTS OF THE LORENTZ AND CPT SYMMETRIES
AT THE PLANCK ENERGY SCALE WITH X-RAY AND GAMMA-RAY OBSERVATIONS}

\author{HENRIC KRAWCZYNSKI, FABIAN KISLAT, MATTHIAS BEILICKE,\\
and ANNA ZAJCZYK}
%\vspace*{-0.05cm}

\address{
Physics Department and McDonnell Center for the Space Sciences\\
Washington University, 
St.\ Louis, MO 63130, USA%, E-mail: krawcz@wuphys.wustl.edu
}

%\vspace*{-0.25cm}

\begin{abstract}
X-ray and gamma-ray observations of astrophysical objects at cosmological distances can be used to
probe the energy dependence of the speed of light with high accuracy and to search for 
violations of Lorentz invariance and CPT symmetry at the Planck energy scale. 
In this conference contribution, we discuss these searches in the theoretical framework of the 
Standard-Model Extension. We present new limits on the dispersion relation governed by 
operators of mass dimension $d=5$ and $d=6$, and we discuss avenues for future progress.
\end{abstract}

\bodymatter

%\vspace*{-0.25cm}

\section{Search for new physics at the Planck energy scale}
% from X-ray and $\gamma$-ray observations of extragalactic sources}

The Standard-Model Extension (SME) provides a theoretical framework 
for studying and constraining Lorentz invariance and CPT symmetry 
violations.\cite{sme1} It is an effective field theory that assumes that our theoretical 
framework --- the Standard Model of particle physics and the theory of General Relativity ---
is the low-energy limit of a more fundamental theory that describes physics at  
higher energies, possibly the Planck energy scale 
($E_{\rm P}= \sqrt{c^5\hbar/G}$ $\approx 1.22 \times 10^{19}$~GeV). 
The action of the Standard Model is considered 
to be the zeroth-order term in an expansion approximating the full theory.  
The SME considers additional terms in the action, and makes it possible to 
use observational data to constrain the magnitude of the additional terms
and thus to quantify the accuracy to which the Standard Model has 
been tested.\cite{tables}
The additional terms are ordered according to the mass dimension $d$ of the operators, and 
operators of dimension $d>4$ lead to a modified photon dispersion relation. 
As the deviation of the group velocity from its low-energy value is expected to scale 
with $(E/E_{\rm P})^{d-4}$ ($E$ being the photon energy), experimental constrains 
on the leading-order terms ($d\,=5, 6$) are the most interesting ones. 
Even though the observations use photons with $E\ll E_{\rm P}$, 
the observations are sensitive to new physics at the Planck energy scale 
as tiny deviations accumulate over large distances.\cite{amelino} 

Gamma-ray burst (GRB) observations with the Fermi $\gamma$-ray telescope have been used
to set sensitive limits on the deviations of the speed of light from SME operators with $d=5$. 
Writing the energy dependence of the
group velocity as $\delta v=\zeta^{(d)0} (E/E_{\rm P})^{d-4}$ (with $\delta v$ in units of $c$) 
the Fermi time-of-flight observations imply $|\zeta^{(5)0}|<0.13$.\cite{data1}  
Here and in what follows, we adopt the notation of Kosteleck\'y \& Mewes\cite{grb2013}, but
suppress the direction dependence of $\zeta^{(d)0}$ . We will do the same below for the 
coefficients $|\zeta^{(d)a}|$ (with $a=3$ or $a=+$) describing 
the birefringent propagation of light. Similar constraints on $|\zeta^{(d)0}|$ follow from AGN 
observations with the H.E.S.S., MAGIC, and VERITAS experiments.\cite{data2} 
These observations are in tension with the expectation $\delta v\sim 1$ 
at $E\approx E_{\rm P}$ (which implies $|\zeta^{(5)0}|\sim 1$). 

In the SME (and in many quantum gravity theories\cite{amelino}) 
the dispersion relation depends not only on the 
photon energy and propagation direction but also the photon helicity. 
For $d=5$ ($\delta v\propto E$), {\it time dispersion and birefringence are governed 
by the same SME coefficients},  and polarization observations constrain the 
coefficients much more sensitively than time-of-flight measurements. Denoting the birefringent 
modifications of the group velocity governed by operators of dimension $d$ as 
$\delta v=\zeta^{(d)a} (E/E_{\rm P})^{d-4}$ with $a=3,+$, the observation of polarized 
UV/optical emission from the afterglows of two GRBs implies $\zeta^{(5)a}<2\times10^{-7}$.\cite{fan}
The detection of polarized GRB emission in the $\gamma$-ray band can give even more 
sensitive constraints, but the significance of the detections so 
far is still marginal.\cite{grb2013}

%\vspace*{-0.25cm}
\section{Constraining $d=6$ coefficients with $\gamma$-ray observations}
% of AGNs and GRBs}

The strong suppression of the corrections to the dispersion relation from operators with $d=5$ may 
find its natural explanation in the fact that these operators break CPT symmetry. 
The leading-order corrections may be caused by the CPT-conserving operators of mass dimension $d=6$. 
Also for the case $d=6$, polarimetric observations give more sensitive constraints than time-of-flight measurements. 
However, in the case of $d=6$ the phenomena of time dispersion and birefringence
are not governed by the same SME operators, justifying a dedicated search for energy dependent time delays 
proportional to $E^2$. The function  $\zeta^{(6)0}$ can be expanded into 25 spherical harmonics.\cite{sme1} 
The 25 complex expansion coefficients satisfy a reality condition and are thus given by 25 real coefficients.
At the conference, rigid constraints on the 25 real coefficients based on the analysis
of 25 sources were presented for the first time. The analysis uses published results of four GRBs detected with Fermi\cite{data1}, one GRB detected 
with RHESSI, and three AGNs detected with the $\gamma$-ray telescopes H.E.S.S., MAGIC, 
and Whipple.\cite{data2}  We derive additional constraints from the publicly available light curves of 
17 Fermi AGNs in two energy bands (300 MeV-1 GeV and 1 GeV-300 GeV).
The analysis uses a simple discrete correlation function analysis to constrain the 
time lag between the two energy bands. 

Figure 1 shows the distribution of the 25 sources in the sky
and the derived 95\% upper limits on the $\zeta^{(6)0}$ (lower limits were derived as well, but are not shown here). 
The most sensitive constraints come from the Cherenkov telescope observations of AGNs
and from the Fermi observation of a bright GRB. Based on these results lower 
and upper limits on the 25 real coefficients describing the direction dependence of $\zeta^{(6)0}$ were derived.
Presently, these latter limits are heavily impacted by the Fermi AGN limits and 
are of the order of $\rm 10^{27}$. Work is underway to improve on the Fermi 
AGN limits with a dispersion cancellation method that uses the full information about
the energy and arrival times of the detected photons.\cite{scargle} 
Note that the results do not yet exclude order-unity modifications of the dispersion relation 
at the Planck energy scale. The same holds true for the constraints from the 
UV/optical polarimetry observations mentioned in the previous section which fall 
short by more than 18 orders of magnitude.   

\begin{figure}[t]
\begin{center}
%\hspace*{-1cm}
\begin{minipage}{1.75in}
\vspace*{-1cm}
\psfig{file=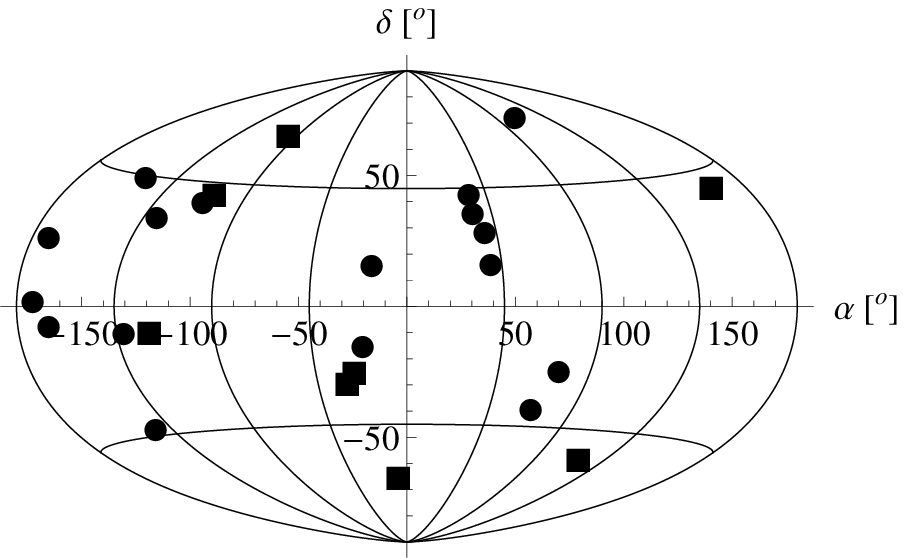,width=2in}
\end{minipage}
\hskip 0.3in
\begin{minipage}{2.4in}
\psfig{file=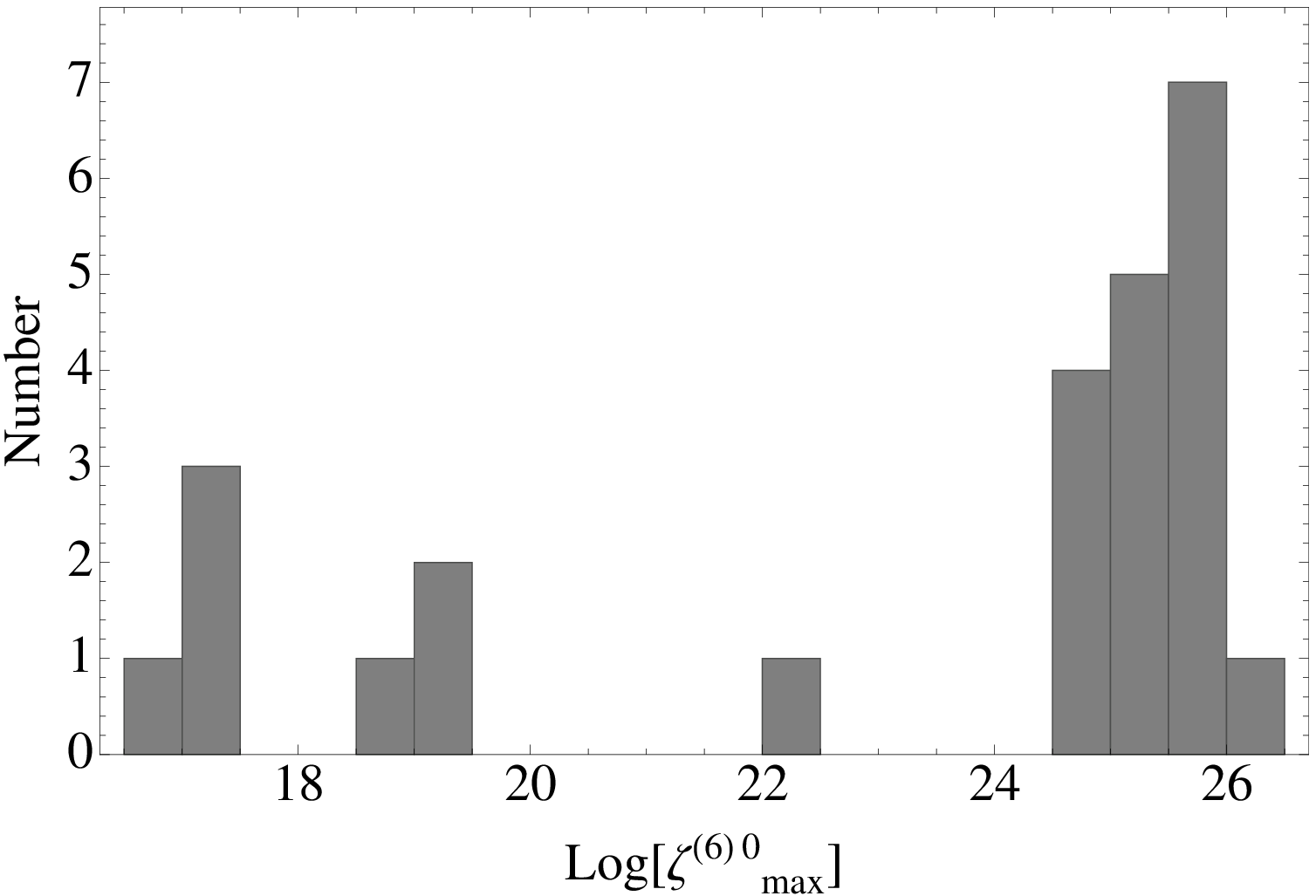,width=2.25in}
\end{minipage}
\end{center}
%\caption{Exemplary results from the analysis discussed in the text: distribution of he sources in the sky (left panel, circles: Fermi AGNs, squares: other sources) and 95\% confidence level upper limits on $\zeta^{(6)0}$. The corresponding lower limits are not shown.}
\caption{Left: sky distribution of the sources 
(circles: Fermi AGNs, squares: other sources).
Right: 95\% confidence level upper limits on $\zeta^{(6)0}$.}
\label{fig1}
\vspace*{-0.1cm}
\end{figure}

\begin{figure}
\begin{center}
\psfig{file=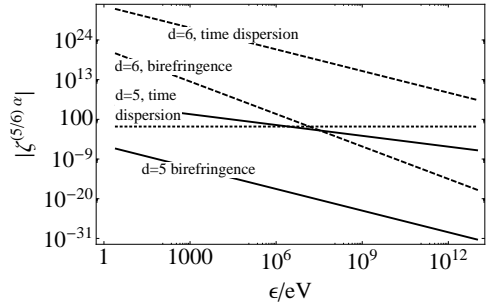,width=2.5in}
%\begin{minipage}{3.5in}
%\hspace*{-0.1cm}
%%\psfig{file=future.eps,width=3.25in}
%\end{minipage}
%\hspace*{-0.24in}
%\begin{minipage}{1.15in}
%\vspace*{-0.5cm}
%\end{minipage}
\end{center}
\caption{Estimates of the limits on $\zeta^{(5)\alpha}$ and $\zeta^{(6)\alpha}$ ($\alpha=0,3,+$) 
that can be derived from future time dispersion and birefringence observations % \protect\newline
 at energy $\varepsilon$. 
Results at and below the dotted line ($\zeta^{(5/6)\alpha}=1$) constrain effects at the Planck energy scale.    }
\label{fig2}
%\vspace*{-0.5cm}
\end{figure}

%\vspace*{-0.25cm}
\section{Conclusions and outlook}

As mentioned before, polarimetric observations rule out a modification of the
photon dispersion relation of order unity at the Planck-scale from operators with $d=5$ by more than six orders of magnitude. 
In contrast, neither time-of-flight measurements nor polarimetric observations do so for the case of  $d=6$. 
It is instructive to evaluate how much better future time-lag and polarization measurements will do in this regard.
For this purpose we assume that observations of GRBs at $z=1$ can constrain the time-of-flight difference 
of photons of energies $E_1=E$ and $E_2=0.1 E$ with an accuracy of 1 msec, and succeed to detect 
a polarized signal from these GRBs. Figure 2 shows the resulting constraints. 
Interestingly, the time-of-flight measurements will not have the 
sensitivity required to constrain new physics at the Planck scale for the case of $d=6$. Polarization observation do better, 
but require the detection of polarized signals at $>$20 MeV energies. Such detections might be possible 
with a next-generation Compton telescope. 

%\vspace*{-0.25cm}


\begin{thebibliography}{x}

\bibitem{sme1}
D. Colladay and V.A. Kosteleck\'y, Phys.\ Rev.\ D {\bf 55}, 6760 (1997); 
Phys.\ Rev.\ D {\bf 58}, 116002 (1998); 
V.A.\ Kosteleck\'y, Phys.\ Rev.\ D {\bf 69}, 105009 (2004);
V.A.\ Kosteleck\'y and M.\ Mewes, Phys.\ Rev.\ D {\bf 80}, 015020 (2009).

\bibitem{tables}
{\it Data Tables for Lorentz and CPT Violation,}
V.A.\ Kosteleck\'y and N.\ Russell,
2013 edition, arXiv:0801.0287v6.

\bibitem{amelino}
G.\ Amelino-Camelia \etal, Nature {\bf 393}, 319 (1998); 
L.J.\ Garay, Phys.\ Rev.\ Lett.\ {\bf 80}, 2508 (1998);
R.\ Gambini and G.\ Pullin, Phys.\ Rev.\ D {\bf 59}, 124021 (1999);
D.\ Mattingly, Liv.\ Rev.\ Rel.\ {\bf 8}, 5 (2005);
T.\ Jacobson, S.\ Liberati, and D.\ Mattingly, 
Ann.\ Phys.\ {\bf 321}, 150 (2006).

\bibitem{data1}
V.\ Vasileiou \etal, Phys.\ Rev.\ D {\bf 87}, 122001 (2013).

\bibitem{grb2013}
V.A.\ Kosteleck\'y and M.\ Mewes, 
Phys.\ Rev.\ Lett.\ {\bf 110}, 201601 (2013).

\bibitem{data2}
S.E.\ Boggs \etal, 
Ap.\ J.\ {\bf 611}, L77 (2004);
A.\ Abramowski \etal, Astrop.\ Phys.\ {\bf 34}, 738 (2011);
J.\ Albert \etal, Phys.\ Lett.\ B {\bf 668}, 253 (2008);
S.D.\ Biller \etal, Phys.\ Rev.\ Lett.\ {\bf 83(11)}, 2108 (1999).

\bibitem{fan}
Y.Z.\ Fan, D.M.\ Wei, D.\ Xu, MNRAS {\bf 376}, 1875 (2007).

\bibitem{scargle}
J.D.\ Scargle, J.P.\ Norris, J.T.\ Bonnell, ApJ, {\bf 673}, 972 (2008).

\end{thebibliography}
\end{document}